\documentclass[twocolumn,apl,amsmath,amssymb,showpacs]{revtex4-2}
\usepackage{epsf}      
\usepackage{graphicx}
\usepackage{color}
\usepackage{soul}
\usepackage{gensymb}
\usepackage{sidecap}
\usepackage{amsmath}
\usepackage{mathtools}
 \usepackage{multirow}
\usepackage[hidelinks,colorlinks=true,linkcolor=blue,citecolor=blue]{hyperref}
\DeclareUnicodeCharacter{2212}{\textendash}

\begin{document}
\title{Magnetic correlations and Griffith-like phase in Co$_2$TiSi$_{0.5}$Al$_{0.5}$ Heusler alloy}

\author{Priyanka Yadav}
\affiliation{Department of Physics, Indian Institute of Technology Delhi, Hauz Khas, New Delhi-110016, India}
\author{Brajesh K. Mani}
\email{bkmani@physics.iitd.ac.in}
\affiliation{Department of Physics, Indian Institute of Technology Delhi, Hauz Khas, New Delhi-110016, India}
\author{Rajendra S. Dhaka}
\email{rsdhaka@physics.iitd.ac.in}
\affiliation{Department of Physics, Indian Institute of Technology Delhi, Hauz Khas, New Delhi-110016, India}

\date{\today}      

\begin{abstract}
We present a comprehensive study aimed at elucidating the complex magnetic correlations in Co$_2$TiSi$_{0.5}$Al$_{0.5}$ Heusler alloy  having the partial B2-type structure amid L2$_1$ cubic main phase. The thermo-magnetization measurements at 100 Oe reveal the presence of two magnetic transitions at T$\rm_{C1}=278~K$ and T$\rm_{C2}=270~K$, respectively, with saturation magnetization of around 1.2 $\mu \rm_ B$/f.u. at 5~K. Our magnetic field dependent studies reveal the dominance of T$\rm_{C1}$ transition at lower fields ($\mu_0\rm H \leqslant 0.03~Tesla$); however, at higher fields the T$\rm_{C2}$ transition becomes more pronounced. The observation of remnant magnetization above Curie temperature suggests the development of Griffiths-like phase, which is extensively analyzed through {\it ac} and {\it dc}- magnetic susceptibility ($\chi$) data. The observed frequency dispersion in imaginary component of $ac-\chi$ indicates blocking of spins leading to the development of distinct magnetic state even below Griffiths-like phase. The evaluation of magnetocaloric potential indicates second order phase transition with notable $\Delta S_{\rm M}=$ 2.22 Jkg$^{-1}$K$^{-1}$ at 7 Tesla. The low-field ($\mu_0\rm H \leqslant 0.1~Tesla$) magnetic entropy ($\Delta S_{\rm M}$) curves exhibit two non-identical positive peaks, which corroborate the existence of T$\rm_{C1}$ and T$\rm_{C2}$ transitions. The nature and range of spin interactions near T$\rm_C$ were scrutinized through rigorous critical phenomenon analysis, and the values of exponents $\alpha, \beta, \gamma$ and $\delta$ to be $0.063, 0.361, 1.108$ and $3.943$, respectively, which are found to be slightly deviating from mean-field theory towards 3D Heisenberg model. Additionally, the exchange magnetic interactions are found to decay as $J(r) \sim r^{-4.6}$. Furthermore, the density functional theory results reveal the half-metallic nature exhibiting 100\% spin polarization (SP). However, the electronic and magnetic properties are greatly affected by the incorporation of structural disorder, which results in drastic reduction of SP to mere 8.3\%.  
\end{abstract}
\maketitle

\section{\noindent ~Introduction}

Heusler alloys (HA) represent an intriguing class of materials known for their diverse physical properties and prospective applications in domains like, spintronics, thermoelectrics, magnetocalorics, spin torque transfer, and topological properties \cite{Zutic_RMP_2004, Galanakis_PRB_2002, Krenke_Nat_2005, Groot_PRL_1983, Kubota_APL_2009, SchulzPRB24}. However, it is well proclaimed that HAs are susceptible to structural disorder (SD), which significantly influences their electronic band structure and magnetic properties. There are numerous studies in literature which have mapped the origin of exotic phenomena, like giant magneto-resistance, exchange-bias effect, large anomalous Hall effect, and giant magneto-caloric effect in HAs to inherent SD \cite{Gercsi_JPCM_2007, Okamura_JAP_2004, Geiersbach_JMMM_2002, Okamura_APL_2005}. In this line, significant tunnelling magnetoresistance (TMR) has been reported in systems like Co$_2$(Cr$_{0.6}$Fe$_{0.4}$)Al/AlO$_x$/CoFe/NiFe/IrMn/Ta heterostructure, which exhibits 16$\%$ TMR at room temperature (RT) and 26.5$\%$ at 5~K \cite{Elmers_PRB_2003}, whereas, the A2-ordered Co$_2$FeAl shows 47$\%$ TMR at RT \cite{Okamura_APL_2005}. These substantial results are attributed to the presence of disorder and/or interfaces within the structures, highlighting the potential of these materials for spintronics applications \cite{Inomata_JJAP_2003, Kubota_JJAP_2004}. Additionally, Block {\em et al.} reported a large magnetoresistive effect (30$\%$ at RT) in B2-ordered Co$_2$(Cr$_{0.6}$Fe$_{0.4}$)Al system \cite{Elmers_PRB_2003}. Furthermore, the low damping constant associated with HAs enhances the current induced dynamics of skyrmions \cite{Mizukami_JAP_2009}. Conversely, the inherent SD appears to influence the shape and size of developed skyrmions through the pinning effect \cite{Gross_PRM_2018}.

Another alluring magnetic state correlated with SD is the emergence of highly inhomogeneous magnetic region, often identified as Griffiths-like phase (GP) \cite{Karmakar_JPCM_2013, Griffith_PRL_1969, Ouyang_PRB_2006, Magen_PRL_2006}. The development of GP in systems such as diluted ferromagnetic maganites, oxides, intermetallics, and semiconductors is often attributed to the presence of quenched disorders and/or competing magnetic interactions \cite{Ouyang_PRB_2006, Pal_APL_2019, Silva_JPCM_2020, Karmakar_JPCM_2013}. For instance, in Pr$_2$CoFeO$_6$, random distribution of Co and Fe ions introduces AFM/FM competing interactions \cite{Pal_APL_2019} while dilution by non-magnetic Co$^{3+}$ ion in La$_{1.5}$Sr$_{0.5}$CoMn$_{0.5}$Fe$_{0.5}$O$_6$ leads to the loss of long range magnetic ordering \cite{Silva_JPCM_2020}, thus giving rise to GP in these systems. Interestingly, the observed GP state in Fe$_2$V$_{1-x}$Rh$_x$Ga HA was attributed to the dilution of Fe atoms upto a critical concentration of $x\leqslant$0.15, hence resulting in development of isolated FM clusters near T$\rm_C$ \cite{Nag_PRB_2023}. In Fe$_{2-x}$Mn$_x$CrAl system, the observed (Fe-Al) anti-site disorder was considered responsible for the evolution of short-range FM clusters at higher temperatures  \cite{Kavita_SR_2019}. More recently, the observed GP state in CoFeVAl system spans an unusually large temperature range, which is also evident in subsequent transport measurements, where the reduced electron-magnon (e-m) scattering resulted in negligible MR signal even under high fields \cite{B_PRB_2024}. Moreover, systems with GP singularity have been observed to display colossal magnetoresistance and giant magnetocaloric effect \cite{Zhang_JAP_2022, Salamon_PRL_2002}. In the context of magnetocaloric applications, HA have garnered much attention since Liu {\it et al.} reported evidences of giant MCE in Ni-(Co)-Mn-In, with $\Delta$T$\rm_{ad}$ = - 6.2 K at 1.9 Tesla, which was comparable to benchmark Gd-based alloys \cite{Liu_NatMat_2012}. The exploration of HAs as magnetocaloric materials opens new avenues for the development of eco-friendly and energy-efficient refrigeration systems, cooling units and heat pumps \cite{Nehla_JAP_2019}. 

In this context, the Co$_2$YZ HAs are of particular interest due to their high Curie temperature T$\rm_C$ and the variation in magnetic moment at Co site, ranging from 0.3 to 1 $\mu_B$, which is strongly influenced by the local environment \cite{Nehla_JAP_2019, GuptJVSTA23}. The Co$_2$TiZ series, in particular, has garnered significant scientific interest for its remarkable properties, including half-metallicity, high spin polarization, and large anomalous Hall effect, highlighting its promising spintronic functionality \cite{Kandpal_JP_2007, Sharma_JMMM_2010, Jezierski_JMMM_1996, Zareii_PBCM_2012}. The variation in magnetic moment at the Co site is also significantly influenced by the localization of 3{\it d} states in constituent transition metals, which is considered as the primary reason for different $\langle L_Z \rangle$/$\langle S_Z \rangle$ contributions at Co site in Co$_2$ZrSn and Co$_2$TiSn, with the Co-3{\it d} states in Co$_2$ZrSn being relatively more localized \cite{Okutani_JPSJ_2000}. Moreover, the theoretically predicted half-metallic characteristics of Co$_2$MSn (M=Ti, Zr, Hf) in \cite{Kandpal_JPDAP_2007, Fecher_JPDAP_2007} are found to be controversial, as few other reports have argued that the presence of Co-{\it d} states near Fermi level of minority spin channel destroys the half-metallicity in some of these materials \cite{Slebarski_PRB_1998, Yamasaki_PRB_2002}. The spin-resolved electronic structure analysis reveals that the majority spin channel exhibits metallic character due to Co$-$Ti interactions. However, direct hybridization between Co$-$Co second next neighbors results in a finite band gap at Fermi level for the minority spin channel \cite{Okutani_JPSJ_2000, Aguayo_JMMM_2011}. Furthermore, the anomalous Hall effect (AHE) in Co$_2$TiAl revealed that an extrinsic mechanism is responsible for the profound AHE signal here, even-though the (e-m) scattering introduces a side-jump contribution, but still skew scattering is observed as the dominant mechanism. Additionally, the observation of large, non-saturating negative MR is also attributed to the significant (e-m) scattering at low fields \cite{Jena_JPCM_2020}. However, to the best of our knowledge, the magnetocaloric properties of Co$_2$TiSi(Al) system have not been extensively explored, except in \cite{Datta_PSSB_2020}. 
 
Therefore, in this paper we present a detailed investigation of the structural and magnetic properties as well as electronic structure of Co$_2$TiSi$_{0.5}$Al$_{0.5}$ (CTSA) Heusler alloy, utilizing the x-ray diffraction, temperature and field-dependent magnetization, and ac susceptibility measurements. Our primary objective is to unravel the origin of unique Griffiths-like phase and the underlying magnetic correlations. It is crucial to thoroughly investigate the intriguing complexities arising from substitution and structural disorder along with their profound influence on the magnetic characteristics. The synergy between our experimental investigations and density functional theory (DFT) simulations aims to bridge the gap between macroscopic observations and understanding microscopic properties of Heusler Alloys.

\section{\noindent ~Experimental and Computational Details}

The polycrystalline Co$_{2}$TiSi$_{0.5}$Al$_{0.5}$ sample was prepared by arc melting technique (CENTORR Vacuum, USA) using water cooled Cu hearth. The constituent elements (Co, Ti, Si, Al) from Alfa Aesar and Sigma Aldrich with purity 99.99+\% taken in the stoichiometric amount were melted in flowing argon environment. To ensure homogeneity of sample, the ingot was turned over and re-melted several times. The ingot wrapped in molybdenum foil was then annealed under vacuum in quartz tube at 1123 K for 7 days in Nabertherm furnace, and we recorded the total weight loss of $<$1\%. 

The x-ray diffraction (XRD) pattern was recorded using Malvern PANalytical conventional diffractometer with Cu K$_{\alpha}$ radiation ($\lambda = 1.5406$\AA) to study the crystallographic structure. The morphology and elemental composition were characterized using field emission scanning electron microscope (FE-SEM) and energy dispersive x-ray (EDX) measurements (see Fig.~1 in \cite{SI}). The magnetic measurements were recorded using MPMS3 SQUID magnetometer from Quantum Design, USA.

The first-principles calculations were performed using the DFT as implemented in the {\em Vienna ab initio simulation package} (VASP) \cite{Kresse_PRB_96} employing a projector augmented wave (PAW) plane wave basis \cite{Kresse_PRB_99}. For the exchange-correlation, we used Perdew, Burke and Ernzerhof (PBE) functional within the generalized-gradient-approximation framework. Further, to obtain the equilibrium structure for ground state, the unit cell was 
optimized using full relaxation calculation with energy convergence and 
atomic force tolerance used as 10$^{-8}$ eV and 10$^{-7}$ eV/\AA, respectively. The self-consistent-field calculations were executed using a k-mesh of 12$\times$12$\times$12 along-with integration method having 0.05 eV smearing and 500 eV plane wave energy cut off.

\section{\noindent ~Results and discussion}

\subsection{Crystal Structure and Magnetic Properties}

\begin{figure}
\includegraphics[width=3.4in]{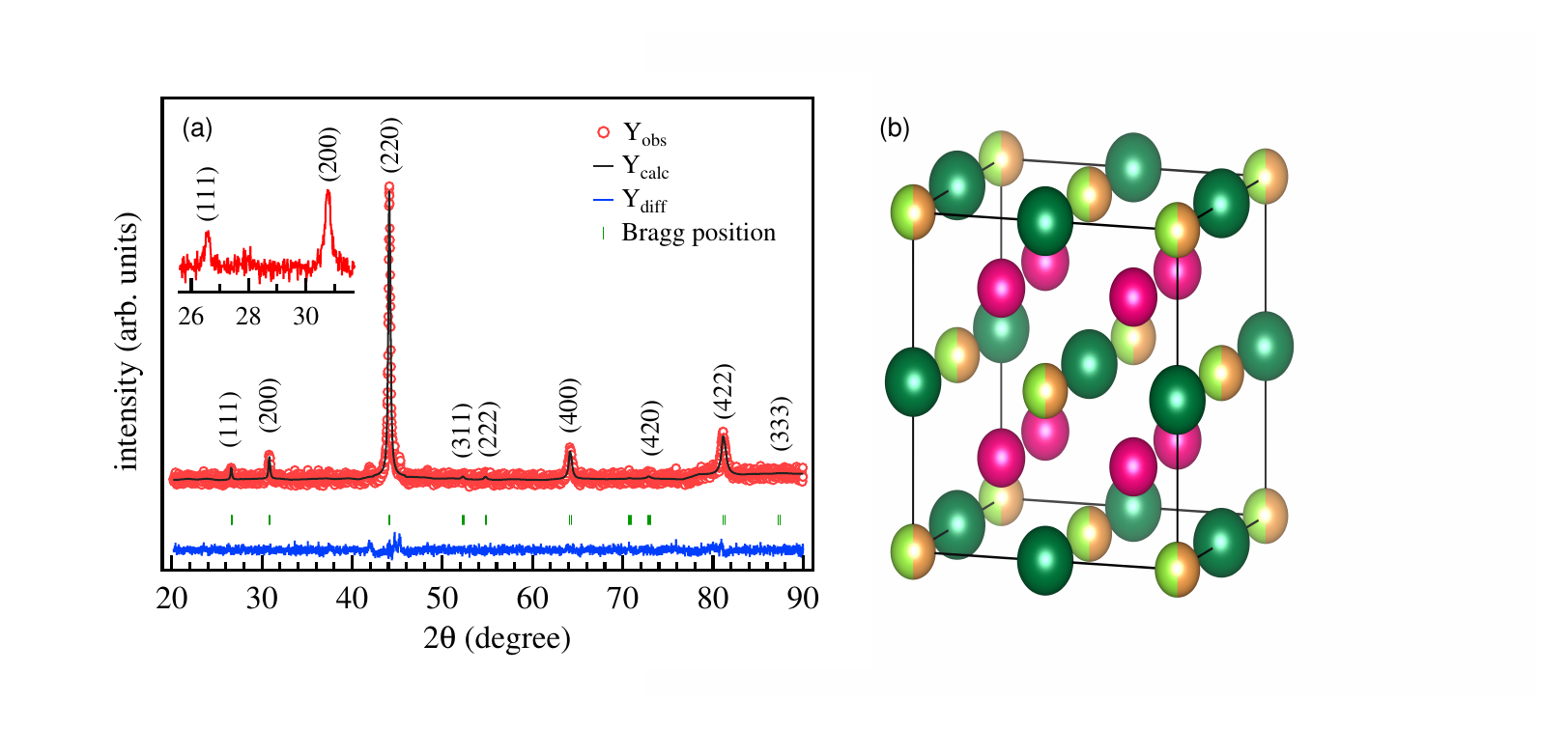}
\caption {(a) The XRD pattern for CTSA sample recorded at room temperature, with the observed data points, simulated curve, difference in observed and simulated data and the Bragg positions, denoted by open red circles, solid black line, solid blue line, and vertical green markers, respectively. The inset highlights the zoomed view of (111) and (200) peaks. (b) The crystal structure of CTSA, crystallized in Fm$\overline{3}$m space group, with Co (pink), Ti (dark green), Si (orange), Al (light green) atoms occupying the (1/4, 1/4, 1/4), (1/2, 1/2, 1/2), (0,0,0) and (0,0,0) sites, respectively.}
 \label{XRD}
\end{figure}

The XRD pattern recorded at room temperature is shown in Fig.~\ref{XRD}(a) where the inset highlights an additional narrow scan of (111) and (200) peaks at 26.7\degree and 30.9\degree, respectively. The Rietveld refinement using Fullprof package is performed considering Fm$\overline{3}$m space group (no. 225) and the extracted lattice parameter is $a=$ 5.8038 \AA, with corresponding fitting parameters as R$_P$ = 2.2, R$_{exp}$ = 2.48 and $\chi^2$ = 1.28. The Bragg reflections corresponding to $h+k+l=4n$, are independent of the state-of-order and are called primary reflections. The supper-lattice reflections, which are order-dependent, can be classified into two sets, $h, k, l$ are odd or $h+k+l=4n+2$. The observation of (111) and (200) peaks [see the inset of Fig. 1(a) for clarity] indicates the presence of ordered L2$_1$ structure; however, the small difference between the atomic scattering factors of Co and Ti, and absorption corrections, render only a qualitative estimate of the chemical order from x-ray diffraction pattern \cite{Webster_JPCM_1973, GuptJAP23}. In Co$_2$TiZ samples the commonly observed preferential disorder is B2-type between the Y(Ti) and Z atoms, hence in the present case, the relatively similar electro-negativities and atomic radii of Ti (1.54, 147 pm) and Al (1.61, 143 pm) atoms in contrast with Si (1.9, 111 pm), leads to favorable cross-site occupation of Ti and Al atomic positions \cite{Felser_SIP}. However, the quantification of the anti-site disorder by refining the respective atomic occupancies at Wyckoff positions could not be concluded as there was hardly any improvement in the refined profiles. Therefore, we try to estimate the degree of disorder from order parameters S and $\alpha$, using the relations S$^2=(\rm I_{200}/I_{220})_{exp}/(I_{200}/I_{220})_{th}$, and S$^2(1- 2 \alpha)^2=(\rm I_{111}/I_{220})_{exp}/(I_{111}/ I_{220})_{th}$, respectively \cite{Johnston_JPCS_1968,Webster_JPCM_1973}. The extracted value of $\alpha$ (fraction of Ti atoms occupying Al sites) from the observed XRD pattern is 0.05 indicating partial B2-type order amid L2$_1$ main cubic phase \cite{Webster_JPCM_1973}. Note that, if the order parameter had been $\alpha$=0.5, it would have symbolized complete order between Ti and Z atoms \cite{Webster_JPCM_1973}. 

\begin{figure} 
\includegraphics[width=3.5in]{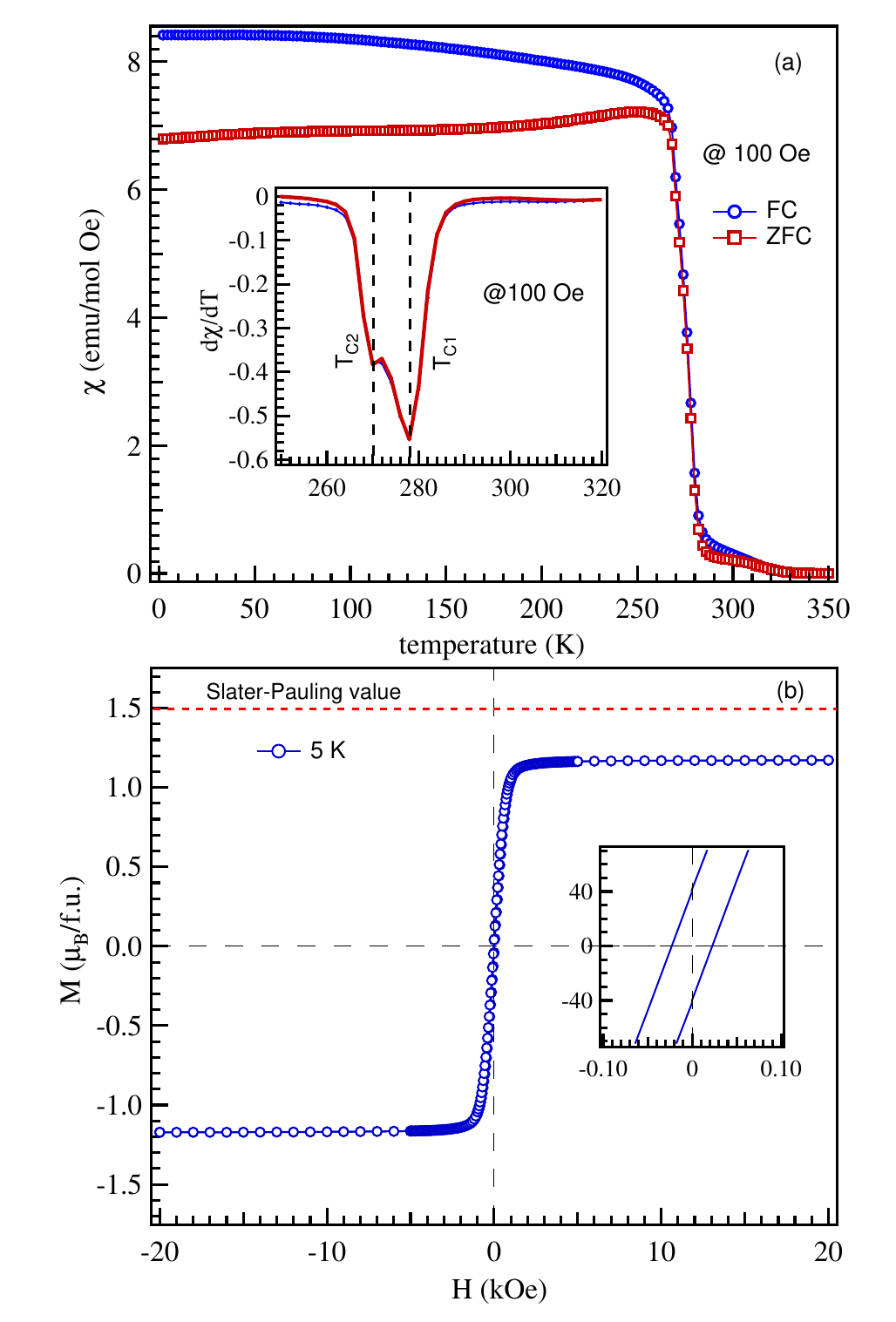}
\caption {(a) The thermo-magnetization curve recorded across the temperature range 2--350 K at 100 Oe magnetic field for ZFC (open red squares) and FC (open blue circles) protocols. The inset illustrates the temperature derivative of susceptibility versus temperature. (b) The isothermal magnetization measured at 5 K, with inset showing the zoomed view emphasizing the soft ferromagnetic character and the red dashed line symbolizes the Slater-Pauling value.}
\label{MT_100}
\end{figure}

\begin{figure}[h]
	\includegraphics[width=3.5in]{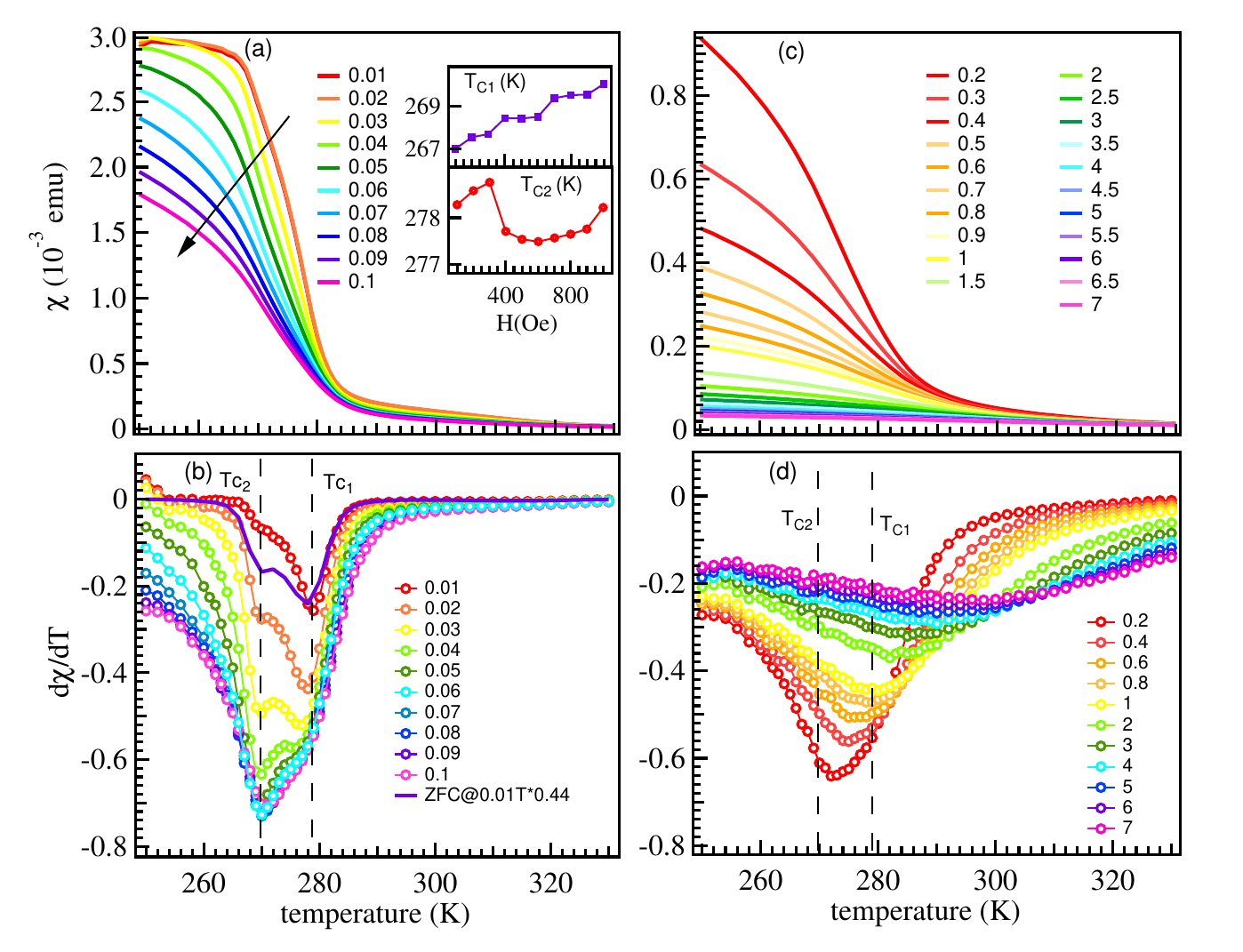}
	\caption{The temperature-dependent magnetic susceptibility curves constructed from magnetic isotherms for temperature range 250--290K corresponding to (a) H $\leqslant$ 0.1 Tesla and (c) 0.1 $\leqslant$ H $\leqslant $ 7 Tesla. (b, d) The first-order temperature derivative of susceptibility plotted as a function of temperature at various values of applied field, H $\leqslant$ 0.1 Tesla and 0.1 $\leqslant$ H $\leqslant $ 7 Tesla, respectively. The solid purple line represents the scaled ZFC recorded data corresponding to 0.01 Tesla applied field, for substantiating the analysis.} 
	\label{MH-virgin}
\end{figure}

In order to investigate the magnetic phase transition and intrinsic magnetic interactions of CTSA, we measure the temperature-dependent magnetization data in zero-field cooled (ZFC) and field-cooled (FC) modes, during warming with an applied field of 100 Oe after cooling the sample in zero field and 100 Oe, respectively. Interestingly, we observe a sharp increase in the magnetic susceptibility ($\chi$) curves below 280~K and as evident in Fig.~\ref{MT_100}(a), the branching of ZFC and FC data points suggests the presence of competing interactions and/or inhomogeneous ferromagnet \cite{Slebarski_PRB_2009}. The bifurcation however is observed to diminish with increasing magnetic field (see Fig.~2 in \cite{SI}). Furthermore, the first-order temperature derivative of $\chi$, clearly shows the presence of two magnetic transitions around $\rm T_{\rm C1}=278~\rm K$ and $\rm T_{\rm C2}=270~\rm K$, indicated by dashed lines in the inset of Fig.~\ref{MT_100}(a). The isothermal magnetization curve recorded at 5~K in Fig.~\ref{MT_100}(b) illustrates low coercivity (around 22 Oe) and the saturation magnetization value is found to be 1.2 $\mu \rm_B$/f.u., which is to some extent below the expected Slater Pauling value of 1.5 $\mu \rm_B$/f.u \cite{Galanakis_PRB_2002, Fecher_JAP_2006}. This slight difference in the magnetic moment values could be ascribed to the presence of atomic disorders in the system \cite{Sokolovskiy_PRB_2012, Baral_JMMM_2019}. The significant influence of structural disorder on magnetic properties has been extensively investigated and discussed in the subsequent DFT section. The temperature-dependent susceptibility ($\chi-\rm T$) curves extracted from the magnetization isotherms are presented in Figs.~\ref{MH-virgin}(a, c) and the corresponding d$\chi$/dT curves are presented in Fig.~\ref{MH-virgin}(b, d), respectively, to illustrate the evolution of the transitions at T$_{C1}$ and T$_{C2}$ with applied magnetic field. The low-field ($\chi-\rm T$) in Fig.~\ref{MH-virgin}(a) (0.01--0.1 Tesla) undergo a notable change in the slope of the curves around 270~K, followed by another transition at about 280~K, which is more evidently observed in the respective d$\chi$/dT plot, Fig.~\ref{MH-virgin}(b) where the transition at T$_{C1}$ appears to be dominant upto 0.03 Tesla; however, further increase in the magnetic field leads to broadening of this feature accompanied with systematic shift of transition temperature towards higher values [see inset Figs.~\ref{MH-virgin}(a)], hence evidencing the ferromagnetic transition. On the other hand, the T$_{C2}$ transition strengthens beyond 0.04 Tesla with a slight dip in the transition temperature at 0.04 Tesla field followed by systematic increase with the field [see inset Figs.~\ref{MH-virgin}(a)], thereby manifesting as another competing ferromagnetic interaction. The possible anti-site disorder between Ti and Al atoms, thereby allocates different exchange-coupling constants between Co--(Ti/Al)--Co sites, which might be associated with the emergence of these magnetic transitions \cite{Reis_JMMM_2017}. Similar observations are reported in DyMn$_2$Si$_2$ where the presence of small defects in crystal structure and/or in-homogeneity resulted in the occurrence of multiple ferromagnetic/anti-ferromagnetic transitions \cite{Reis_JMMM_2017}.   
 
 \begin{figure*}
	\includegraphics[width=7.2in]{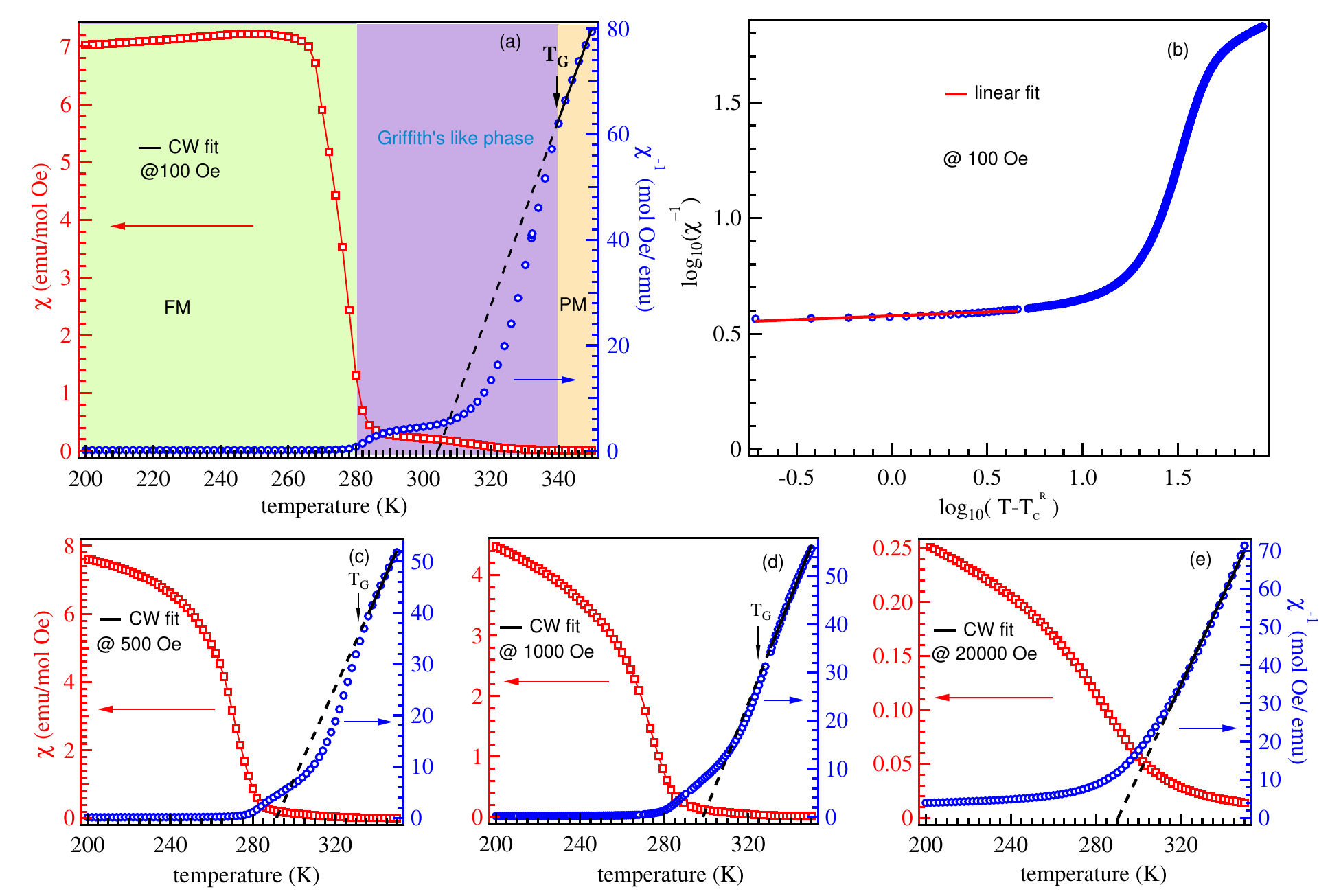}
	\caption {(a, c--e) The thermo-magnetization curves recorded across temperature range 2--350K at 100, 500, 1000, 20000 Oe magnetic fields, $\chi$ (open red squares) and $\chi^{-1}$ (open blue circles). The solid black line represents CW fit in high temperature regime. The arrows denote the respective Griffith temperatures for respective applied field. (b) The $log (\chi^{-1})$ versus $log(T-T_{\rm C}^{\rm R})$ curve representing the Griffith phase singularity at 100 Oe along with linear fit (solid red line).} 
	\label{GP2}
\end{figure*}

To elucidate the fundamental magnetic interactions, the inverse susceptibility ($\chi^{-1}$) is plotted as a function of temperature in Fig.~\ref{GP2}(a). The high-temperature behavior exhibits a slight curvature, often associated with the presence of short-range magnetic correlations and/or temperature-independent susceptibility contribution ($\chi$$_0$) \cite{Mugiraneza_CP_2022}. We fit the susceptibility data in 340--350~K range with Curie Weiss (CW) law, as below:
\begin{equation} 
  \chi= \frac{C}{(T - \theta_{\rm CW})}
  \label{CW}
\end{equation}
where $C$ and $\theta_{\rm CW}$ are the Curie constant, and Curie temperature, respectively. The obtained value of the effective magnetic moment ($\mu \rm_{eff} = 2.13~\mu \rm_B$), and the positive value of $\theta_{\rm CW}=304.3~\rm K$ indicates the ferromagnetic-type interactions in the sample. Interestingly, the $\chi^{-1}$ shows a downward deviation from CW fit and takes a sharp downturn before reaching the Curie temperature. This feature implicitly suggests a Griffiths-like phase (GP) transition, as observed in Fe$_{2-x}$Mn$_x$CrAl \cite{Kavita_SR_2019}, where the anti-site disorder between Fe$-$Al atoms resulted in random allocation of different exchange coupling constants at different lattice sites. This led to the coexistence of competing magnetic correlations hence suppressing the PM ordering by forming small ferromagnetic clusters well above T$_C$ \cite{Chatterjee_JPCM_2022, Nag_PRB_2023, Barnabha_PRB_2024}. The deviation temperature of $\chi^{-1}$ from CW fit, referred to as the Griffith temperature T$_G$, marks the onset of short-range magnetic ordering in the system. Therefore, we examine the field dependence of $\chi-\rm T$, which was found to be similar across different fields, as evident from Figs.~\ref{GP2}(a, c, d, e). However, as the field increases above 100 Oe, the prominent downturn softens and completely vanishes at 2~Tesla, exhibiting CW behavior at high fields. Likewise, Barnabha {\it et al.} observed an identical suppression in the downturn with increasing applied field (up to 2 kOe) in CoFeVAl, due to the dominant paramagnetic contribution at higher fields \cite{Barnabha_PRB_2024}. 

Intriguingly, in the present case, the $\rm T_{\rm G}$ is found to shift towards lower temperatures from 338~K to 328~K, respectively, with increasing the field from 100~Oe to 1000~Oe, indicated with arrows in the Figs.~\ref{GP2}(a, c, d). These features hence align well with the characteristic signatures of Griffiths-like phase, which is observed to emerge due to the nucleation or local alignment of small ferromagnetic clusters within globally paramagnetic regime \cite{Slebarski_PRB_2024, Ulrich_PRB_2003, B_PRB_2024, Nag_PRB_2023}. The magnetic response of a system exhibiting GP phase is governed by the largest cluster or correlated volume \cite{Griffith_PRL_1969,Jiang_JPCM_2009}. Therefore, examining the logarithmic dependence of inverse susceptibility on temperature scale is crucial, where the GP singularity ($\chi^{-1}$) can be characterized by the power 
law \cite{Jiang_JPCM_2009}
\begin{equation} 
	\chi^{-1} \propto (T - T_{\rm C}^{\rm R})^{1 - \lambda},
  \label{lambda}
\end{equation}
here, the $T_{\rm C}^{\rm R}$ and $\lambda$ are the characteristic temperature 
of the random ferromagnet and exponent of the power law, respectively. The 
exponent $\lambda$ symbolizes the deviation from CW nature and lies between 
0$<\lambda<$1 for GP. Meanwhile, the characteristic temperature $T_{\rm C}^{\rm R}$ 
plays a significant role in the aforementioned analysis and lies between the actual and the
highest ordering temperature, $T_{\rm C}<T_{\rm C}^{\rm R}<T_{\rm G}$, 
permissible by exchange bond distribution. In the present analysis, we use 
$T_{\rm C}^{\rm R}=\theta_{\rm CW}$, as it ensures that
equation~(\ref{lambda}) reduces to CW nature in PM region for $\lambda\sim$ 0. 
The $log (\chi^{-1})$ versus $log(T-T_{\rm C}^{\rm R})$ data presented in 
Fig.~\ref{GP2}(b) for 100 Oe demonstrates a small 
linear region at high temperatures followed by a downturn and 
another linear region (fitted solid red line) at lower temperatures. The linear fitting of the low-temperature segment of the curve produces $\lambda=0.867\pm0.027$, embodying the GP regime. The large observed value of $\lambda$ 
corresponds to a greater deviation from the CW behavior, indicating the robustness 
of GP in the present sample. Recently, Nag {\it et. al.} have reported a similar robust GP state with $\lambda=0.88$, corresponding to the presence of quenched disorder and coexisting AFM/FM interactions in CrFeVGa Heusler alloy \cite{Nag_PRB_2023}. So we can conclude that the sharp downturn observed above $T_{\rm C}$ in Fig.~\ref{GP2} could be attributed to the dominant contribution from small ferromagnetically coupled clusters over PM ordering. The crystallographic disorder, here in the form of randomly distributed local B2-type phase, suppresses the long range PM ordering temperature, because of the random allocation of exchange-coupling constants at different lattice sites. This develops an intermediate temperature region where, the FM clusters are dispersed within the paramagnetic matrix, thereby resulting in the downward deviation of $\chi^{-1}$ from the CW fit \cite{Mohapatra_PRB_2007}. 

\begin{figure}[h] 
	\includegraphics[width=3.3in]{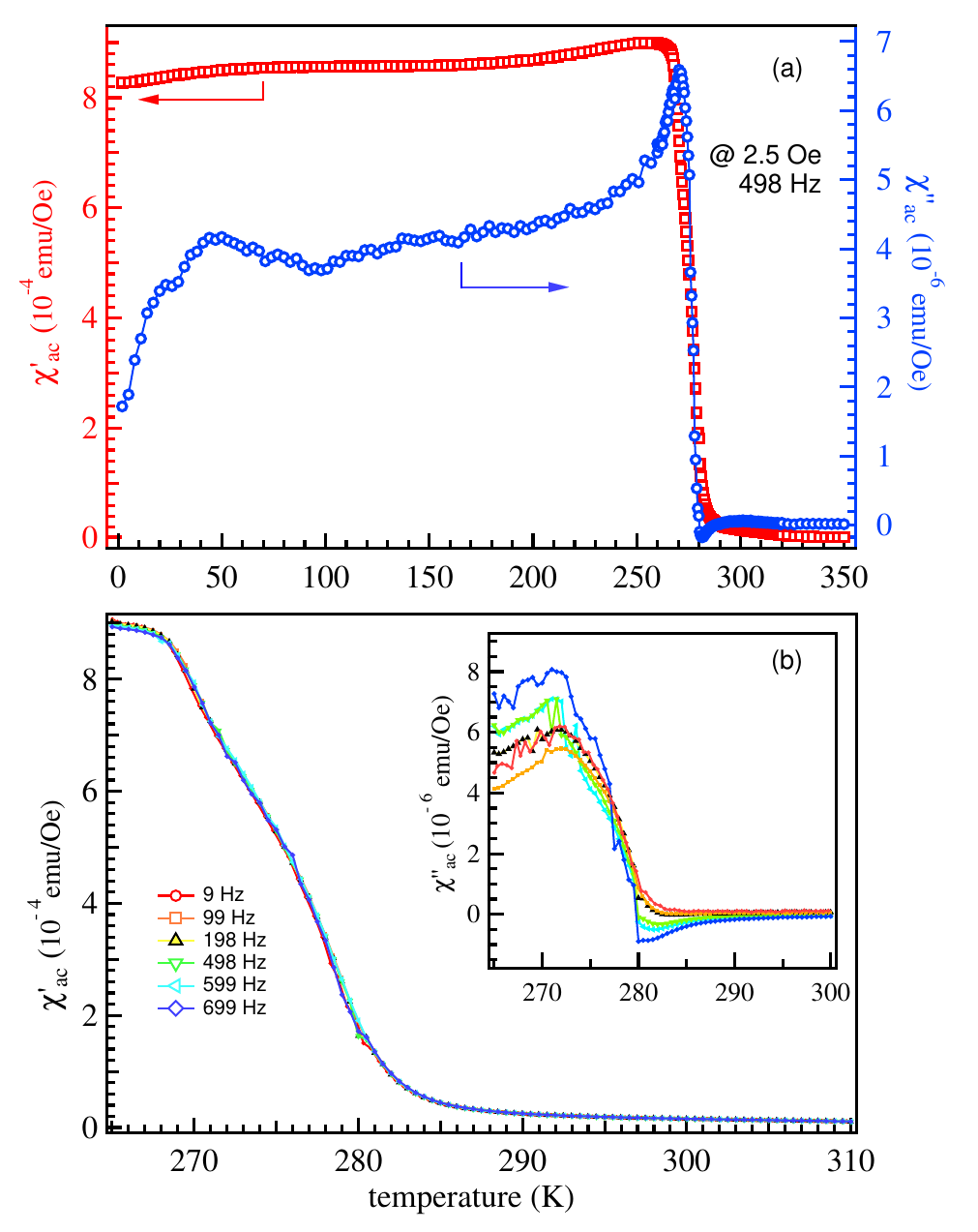}
	\caption {(a) The real 
		$\chi_{ac}'$ (open red squares) and imaginary $\chi_{ac}''$ component (open 
		blue circles) of ac susceptibility measured at 2.5 Oe ac field and 498 Hz exciting 
		frequency. (b) The frequency dependent $\chi_{ac}''$ 
		measured at 2.5 Oe ac field and 9, 99, 198, 498, 599, 699 Hz frequencies in 
		the vicinity of 278 K. } 
	\label{AC}
\end{figure}

Moreover, the {\it ac} magnetic susceptibility ($\chi_{ac}=\chi_{ac}'+ \iota\chi_{ac}''$) 
provides a powerful tool to probe the magnetization and dynamics of a system as a 
function of temperature, frequency, amplitude and direction of {\it ac} and 
bias {\it dc} magnetic fields. Fig.~\ref{AC}(a) presents the data for $\chi_{ac}'$ 
and $\chi_{ac}''$ recorded at 2.5 Oe field and an excitation frequency 
of 498 Hz. Here, the $\chi_{ac}'$ shows a frequency-independent 
peak at about 280~K corresponding to the magnetic phase transition. However, $\chi_{ac}''$ exhibits a low-intensity broad hump at lower temperatures (30--90)K followed by an asymmetric feature near 280~K, see Fig.~\ref{AC}(a). The subsequent frequency-dependent study performed in the vicinity of 280~K reveals an enhancement in the amplitude of this 
feature along with clear frequency dispersion in the $\chi_{ac}''$ component, as shown in Fig.~\ref{AC}(b). This suggests the presence of a distinct magnetic state developed due to blocking of spins below Griffiths phase. Similarly, in Fe$_{2-x}$Mn$_x$CrAl, $\chi''\rm_{ac}$ exhibits frequency dependence near T$\rm_C$, but without shift in the peak position, which was then ascribed to the irreversible domain wall movement and/or pinning effect \cite{Kavita_SR_2019}. Similar frequency dispersion with negative $\chi''_{ac}$ anomaly corresponding to the Griffiths phase has been also observed in Tb$_2$CoMnO$_6$ system \cite{Anand_JMMM_2021}. In several other systems, anomalous relaxation processes have often been attributed to mediate such frequency dependence of the $\chi_{ac}''$ data \cite{Levin_PRB_2000, Higelin_JMMM_1985}. 

\subsection{Magnetocaloric Properties}

\begin{figure*} 
\includegraphics[width=6.9in]{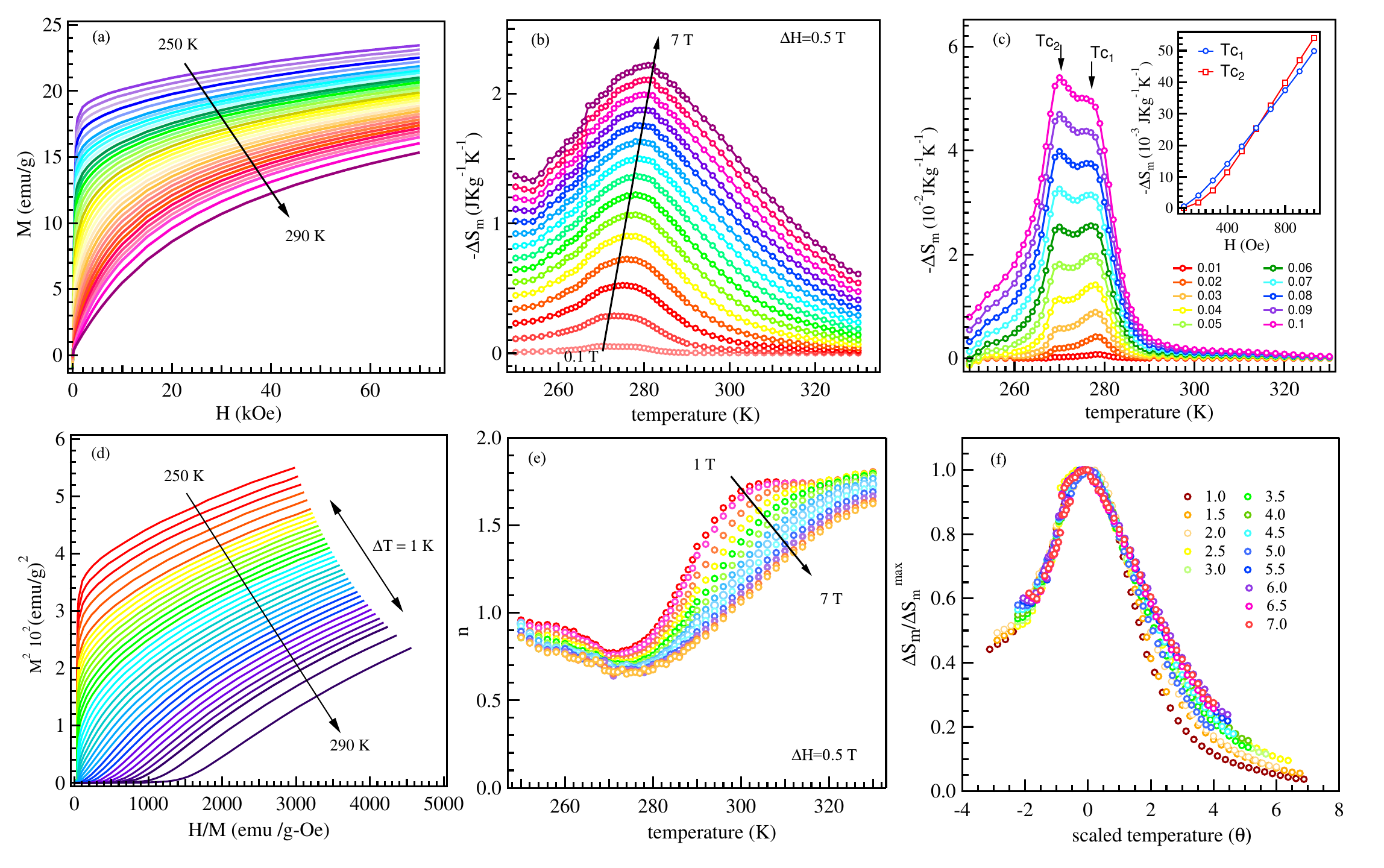}
\caption  {(a) The isothermal magnetization curves for temperature range 250--290 K, upto 7~Tesla applied magnetic field. (b, c) The magnetic entropy change $\Delta S_{\rm M}$ versus temperature plotted as a function of magnetic field (0.1--7 Tesla) and (0.01--0.1 Tesla), respectively. (d) Arrott plot M$^{1/\beta}$ versus (H/M)$^{1/\gamma}$ embodying the mean-field model, is illustrated corresponding to the critical exponents $\beta, \gamma$ as 0.5 and 1, respectively. (e) The variation of exponent "$n$", representing the scaling nature of magnetocaloric properties with applied magnetic fields as a function of temperature. (f) The normalized entropy change plotted against scaled temperature to deduce the order of magnetic phase transition.}
	\label{order}
\end{figure*}

We perform the magnetocaloric analysis to investigate the nexus of these complex magnetic interactions with temperature and magnetic field. The isothermal magnetization measurements were performed near $T_{\rm C}$ (278~K) with a temperature interval of $\Delta$T = 1~K and the magnetic fields upto 7~Tesla, see Fig.~\ref{order}(a). The isothermal entropy change ($\Delta S_{\rm M}$) can be determined using the following Maxwell's equation, 
\begin{equation} 
	\Delta S_{\rm M} =\mu_0 \int_{0}^{H_{max}} \left(\frac{\partial {M(T,H)}}{\partial {T}}\right)_H dH, 
\label{entropy}
\end{equation}
where, the $\mu_0$ represents permeability. Fig.~\ref{order}(b) depicts the obtained entropy change for magnetic fields ranging from 0.1 to 7 Tesla across the $T_{\rm C}$. All the entropy curves exhibit a distinctive peak at temperature $T_{\rm p}$ and its position shifts slightly towards higher temperatures as well as the maximum entropy change increases with increasing magnetic field. For the present sample, the maximum entropy change is found to be 2.22 J.kg$^{-1}$K$^{-1}$, at $\mu_0\rm H =$ 7~Tesla, which is comparable to the related Co-based systems, like 2 J.kg$^{-1}$K$^{-1}$ @ 9~Tesla for the Co$_2$Cr$_{0.75}$Ti$_{0.25}$Al \cite{Nehla_PRB_19} and 2.55 J.kg$^{-1}$K$^{-1}$ @ 5~Tesla for the Co$_{50}$Cr$_{25}$Al$_{25}$ \cite{Panda_JALCOM_2015}. 
Interestingly, in the present sample for lower fields ($\mu_0\rm H \leqslant$ 0.1~Tesla) the entropy curve exhibits two non-identical positive peaks associated with T$_{C1}$ and T$_{C2}$ magnetic transitions, as shown in Fig.~\ref{order}(c), which is consistent with the observations of Fig~\ref{MH-virgin}. The relative magnitude of maximum entropy change $\Delta S_{\rm M}$ in these peaks changes with applied magnetic field, and it instantiates a crossover at about 0.06 Tesla, see the inset of Fig.~\ref{order}(c). Therefore, the development of randomly allocated, temperature and magnetic field dependent exchange interactions in the sample, possibly act as the driving force for the occurrence of the observed magnetic transitions \cite{Levin_PRB_2000,Reis_JMMM_2017}. 

Note that the magnetocaloric properties are strongly influenced by the nature 
of the magnetic transition where the materials having second-order phase transition (SOPT) are desirable for these applications due to the absence of thermal hysteresis, but they render considerably lower $\Delta S_{\rm M}$ and $\Delta T_{\rm ad}$ values.  The order of magnetic phase transition can be determined through magnetic analysis, like the Banerjee criterion, which suggests that for SOPT materials the slope of the Arrott curve for isotherms below and above $T_{\rm C}$ is positive \cite{Banerjee_PL_1964}. Here, the Arrott plot ($M^2$ versus H/M) illustrated in Fig.~\ref{order}(d) indicate the positive slope of the isotherms, affirming the second-order nature of the phase transition. However, this criterion is not always reliable because it assumes that the system adheres to the mean-field model, which may not be the case always. To further 
quantify the nature of the phase transition, we examine the scaling nature of the magnetocaloric properties by probing 
the magnetic field dependence of the entropy change, defined as 
$\Delta S_{\rm M}$ $\propto$ $H^{n}$ \cite{Franco_NatCom_2018, Law_AM_2021}. This general method is not constrained to any particular equation 
of state and has been verified and validated by Law {\it et. al.} for a various
materials, including Heusler alloys such as
Ni$_{45.7}$Mn$_{36.6}$In$_{13.5}$Co$_{4.2}$ \cite{Law_AM_2021}.  
Therefore, we plot the variation of exponent $``n"$ as 
a function of temperature using the following relation \cite{Franco_JMMM_2009}
\begin{equation} 
	{n(T,H)} = \frac{\partial{ ln(|\Delta S_{\rm M}|)}}{\partial{ln (H)}}. 
\label{Maxwell}
\end{equation}
For the SOPT, the value of $n$ should approach 1 below $T_{\rm C}$ and to paramagnetic value 2 for temperatures above $T_{\rm C}$, whereas, any value of $n$ greater than 2 is associated with the presence of first-order phase transition. 
However, before reaching the paramagnetic value, it is expected to exhibit a 
minimum at $T = T_{\rm C}$ \cite{Franco_NatCom_2018}. Our computed data for $n$ as a function of temperature are presented in Fig.~\ref{order}(e), which clearly shows that the $n$ values are  approximately close to 1 for temperatures below $T_{\rm C}$, and then starts approaching a 
minimum at around 278 K before increasing again towards the value of 2. 
This minimum corresponds to the Curie transition from 
ferromagnetic to the paramagnetic regime, which is found to be consistent with the transition 
observed in the thermo-magnetization measurement [as shown in Fig.~\ref{MT_100}(a)]. 

Another qualitative analysis for the SOPT is that after the normalization of entropy 
change and scaling of temperature, all the field-dependent entropy curves should 
collapse into a single curve. This analysis is based on the assumption that equivalent points corresponding to different experimental curves should converge to the same point on the universal curve \cite{CondeFranco_JPCM_2008}. 
The equivalent points, denoted as $T_{\rm r1}$ and $T_{\rm r2}$, are recognized as points corresponding to the same level relative to the peak on the entropy curve. In the present work, for the analysis of 
$\Delta S_{\rm M} (T_{\rm r}) = x \Delta S_{\rm M}^{\rm max}$ we have taken 
the arbitrary factor $x$ as 0.8. It is to mention that, the choice of $x$ 
does not affect the analysis. The temperature is scaled into $\theta$ 
using the following equation \cite{Dong_JAP_2008}
\begin{equation}
   \theta = \left\{
	   \begin{array}{cl}
           -\left(T-T_{\mathrm{pk}}\right)/\left(T_{\mathrm{r} 1}-T_{\mathrm{pk}}\right), & T \leq T_{\mathrm{C}} \\
           \left(T-T_{\mathrm{pk}}\right)/\left(T_{\mathrm{r} 2}-T_{\mathrm{pk}}\right), & T > T_{\mathrm{C}}
   \end{array}\right.,
\end{equation}
where, $T_{\rm r1}$ and $T_{\rm r2}$ represent the temperatures corresponding to $\Delta S_{\rm M} (T_{\rm r1,r2}) = 0.8 ^* \Delta S_{\rm M} (T_{\rm pk})$ 
and $T_{\rm pk}$ is the temperature corresponding to $\Delta S_{\rm M}^{\rm max}$.
Fig.~\ref{order}(e) shows the normalized entropy curves as a function 
of scaled temperatures, which was found to collapse into a single curve, further confirming the SOPT in the present sample. 
Note that the relative cooling power (RCP) is considered as the figure of merit for magneto-caloric materials, which is defined as the product of maximum entropy 
change and temperature at full width half maximum (FWHM) 
$\delta T_{\rm FWHM}$ \cite{Pecharsky_JAP_1999}, as
\begin{equation}
	{\rm RCP} = |-\Delta S_{\rm M}| \times \delta T_{\rm FWHM}.
\label{RCP}
\end{equation}
The value of $\delta T_{\rm FWHM}$ was extracted by fitting the entropy curves in Fig~\ref{order}(b), using the Gaussian function. The highest value of RCP is obtained as 64.4 J.kg$^{-1}$ (for $\delta T_{FWHM}\sim$ 29~K) at $\mu_0\rm H =$ 7~Tesla, which is found to be better than the reported value for Co$_2$TiSi$_{0.75}$Al$_{0.25}$, 55.25 J.kg$^{-1}$ at $\mu_0\rm H=$ 3~Tesla \cite{Datta_PSSB_2020}.

\subsection{Critical Analysis}

To elucidate the microscopic correlations responsible for the intricate magnetic interactions in the present sample Co$_2$TiSi$_{0.5}$Al$_{0.5}$, we now extensively examine the static critical behavior across the FM/PM transition around $T\rm_C=$ 278 K. According to the Landau mean-field theory, the M$^2$ vs H/M curves across the T$\rm _C$ should constitute of parallel straight lines in high-field region, with T = T$\rm_C$ isotherm passing through the origin. However, as shown in Fig.~\ref{order}(d), the Arrott curves are not stringently linear in high-field region, indicating that the system cannot be characterized by mean-field model. Therefore, we use the modified Arrott plot (MAP) approach to characterize the phase transition, i.e., the magnetization isotherms are analyzed using the equation of state proposed by Arrott-Noakes \cite{Arrott_PRL_67}: 
\begin{eqnarray}  
	(H/M)^{1/\gamma}=a\epsilon+bM^{1/\beta},	
	\label{AN}
\end{eqnarray}
where, $a$ and $b$ are temperature-dependent parameters. Near the T$\rm_C$, the spin-spin correlation length ($\xi$), diverges as $\xi=\xi_0|(T-T\rm_C)/T\rm_C|^{-\nu}$, leading to formulation of universal scaling laws governed by asymptotic critical exponents and amplitudes, as \cite{Kouvel_PRL_68}
\begin{eqnarray}  
	M_{S} (0,T)=M_0 (-\epsilon)^\beta   \;\;\; for\; \epsilon<0,   \;T<T_{\rm C} 	
	\label{critical_beta}
\end{eqnarray} 
\begin{eqnarray}  
	\chi_0^{-1} (0,T)=\frac{h_0}{M_0} (\epsilon)^\gamma   \;\;\; for\; \epsilon>0,   \;T>T_{\rm C}	
	\label{critical_gamma}
\end{eqnarray}
\begin{eqnarray}  
	M(H,T_C )=DH^{1/\delta}   \;\;\; for\; \epsilon=0,   \;T=T_{\rm C}			
	\label{critical_delta}
\end{eqnarray}
here, M$_{S}$, $\chi_0^{-1}$ and $\epsilon = (T - T_{\rm C}) / T_{\rm C}$ represent the spontaneous magnetization below T$_{\rm C}$, initial inverse susceptibility above T$_{\rm C}$, and reduced temperature, respectively, while the M$_0$, h$_0$/M$_0$, and D denote the critical amplitudes. 
\begin{figure*}
	\centering
	\includegraphics[width=1
	\textwidth]{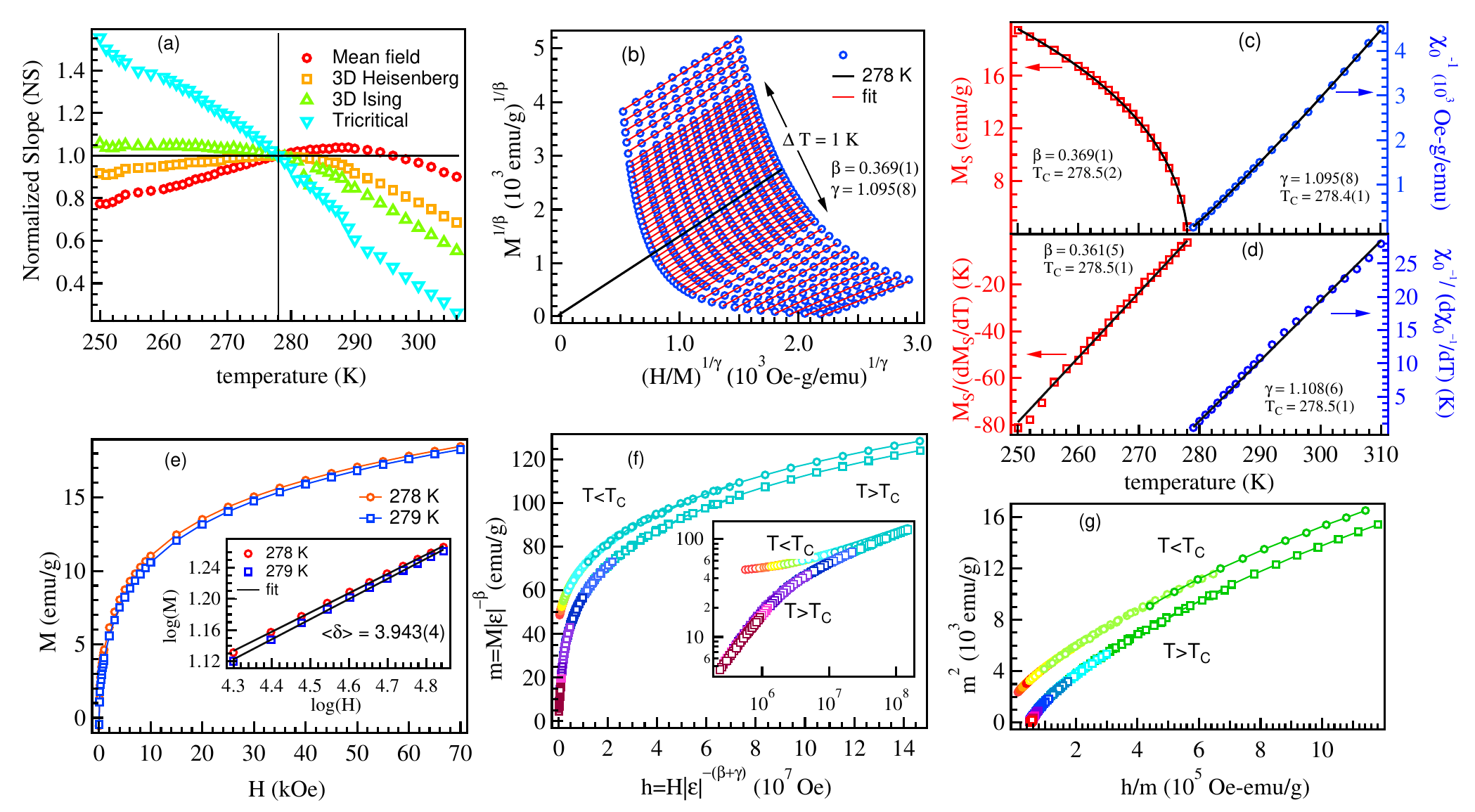}
	\caption {(a) The normalized slopes (NS) for different universality classes in the critical region for Co$_2$TiSi$_{0.5}$Al$_{0.5}$. (b) The modified Arrott plot using the correct critical exponents for the sample, where blue circles, red lines, and black line represent the experimental data points, linear fit for $H \geqslant 2$ Tesla, and the extrapolation of the linear fit at $T_C$ down to $H=0$, respectively. (c) The temperature-dependent spontaneous magnetization $M_S$ (left axis) and inverse initial susceptibility $\chi_0^{-1}$ (right axis), where the black solid curves represent the best fit using equations.~\ref{critical_beta} and \ref{critical_gamma}, respectively. (d) Temperature-dependent $M_{S}(T)/[dM_{S}(T)/dT]$ (left axis) and $\chi_0^{-1}/d\chi_0^{-1}/dT$ (right axis) plots, where the black solid lines represent the linear fit using equations~\ref{KF_beta} and \ref{KF_gamma}, respectively. (e) The $M$-$H$ curves near $T_C$ [278.5(1)~K] at 278 and 279~K, inset shows the log-log plot, where solid black lines represent the linear fit for $H \geqslant 2$ Tesla. (f) The reduced magnetization m versus reduced field h plots, insets show the log-log plot of the same, and (g) the m$^2$ versus h/m plot.
	 }
	\label{critical_1}
\end{figure*}
The MAPs [M$^{1/\beta}$ vs. (H/M)$^{1/\gamma}$] corresponding to critical exponents predicted for the tricritical ($\beta=0.25$ and $\gamma=1$), 3D Heisenberg ($\beta=0.365$ and $\gamma=1.386$), mean-field ($\beta=0.5$ and $\gamma=1$), and 3D Ising ($\beta=0.325$ and $\gamma=1.24$) models \cite{Kaul_JMMM_85}, are constructed and illustrated in Figs.~3(a--d) of \cite{SI}, respectively. The high magnetic field region (H $\geqslant$ 2 Tesla) was then linearly fitted (indicated by the red lines) and we observed that the universality classes exhibit quasi-straight lines over the wide range of fields. To determine which model produces the best fit, we plot the normalized slope (NS)=S(T)/S(T$\rm_C)$, where the slope at any given temperature has been normalized to the slope at T$\rm_C=$  278 K. Fig.~\ref{critical_1}(a) displays the NS extracted from the corresponding MAPs of Figs.~3(a--d) of \cite{SI}. The sample shows least deviation from unity for the mean-field and 3D Heisenberg models as we move away from the critical point (T$_{\rm C}$), indicating that either of these models could successfully characterize the magnetic interactions.

Therefore, we employed an iterative approach to estimate the precise values of the critical exponents $\beta$ and $\gamma$ \cite{Liu_PRB_18, Zhang_PRB_12}. The linear extrapolation of high-field region gives M$_S$(T) and $\chi_0^{-1}$(T) as intercepts on M$^{1/\beta}$ and (H/M)$^{1/\gamma}$ axes, respectively. By substituting these resultants in equations~\ref{critical_beta} and \ref{critical_gamma}, we obtain new set of $\beta$ and $\gamma$ values, which are then used to construct new MAPs. This process is repeated until we are able to find the stable values of $\beta$ and $\gamma$ within the error bars. It is imperative to note that these values of the critical exponents are independent of their initial trial values, thereby confirming their reliability. After rigorous iterations, a set of parallel straight lines were generated corresponding to $\beta$, $\gamma$, and T$_{\rm C}$ as 0.369(1), 1.095(8), and 278.5(1) K, respectively, and presented in Fig.~\ref{critical_1}(b). The isotherm at 278 K passes through the origin [see the black line in Fig.~\ref{critical_1}(b)], thus representing the T$\rm_C$. However, a slight curvature in the low magnetic field region of the isotherms (not shown here) could be attributed to the complex multi-domain spin structure in this regime \cite{Nehla_PRB_19}. The temperature dependence of final M$_{\rm S}$(T) and $\chi_0^{-1}$(T) values acquired from the MAP is plotted in Fig.~\ref{critical_1}(c). To further substantiate the critical exponents and T$\rm_C$, we investigate the M$_S$(T) and $\chi_0^{-1}(T)$ data using Kouvel-Fisher (KF) plot \cite{Kouvel_PRL_68},
\begin{eqnarray}  
	\frac{M_{S} (T)}{dM_{S} (T)/dT}=(T-T_C)/\beta
	\label{KF_beta}
\end{eqnarray} 
\begin{eqnarray}  
	\frac{\chi_0^{-1}}{d\chi_0^{-1}/dT}=(T-T_C)/\gamma
	\label{KF_gamma}
\end{eqnarray} 
According to the KF method, the M$_{\rm S}$(T)/[dM$_{\rm S}$(T)/dT] and $\chi_0^{-1}$/[d$\chi_0^{-1}$/dT] yield linear temperature dependence with slopes of 1/$\beta$ and 1/$\gamma$, respectively, as presented in Fig.~\ref{critical_1}(d). The obtained values of $\beta$, $\gamma$, and T$_{\rm C}$, from the slopes and y-intercepts of linear fitted curves, are 0.361(5), 1.108(6), and 278.5(1) K, respectively, which are consistent with those generated from the MAPs, hence validating their accuracy.

Furthermore, the critical isotherm at T$_{\rm C}$ is characterized by the exponent $\delta$. It exhibits a linear behavior on a log-log scale according to equation~\ref{critical_delta} with slope 1/$\delta$. To verify the value of T$_{\rm C}$ [278.5(1) K] extracted from the MAP and KF methods, we perform the critical isotherm analysis at the magnetic ordering temperature in Fig. \ref{critical_1}(e). The isotherms in asymptotic region were recorded in 1 K steps, therefore, the M-H curves corresponding to 278 K and 279 K are presented in Fig.\ref{critical_1}(e). The linear fitting in the inset of Fig.~\ref{critical_1}(e) gives an average $\delta$=3.943(4). Similar analysis to extract the critical parameters has been reported for few Co-based Heusler alloys, as listed in Table~\ref{Tt}. The critical exponent $\delta$ can also be calculated through the Widom scaling relation \cite{Widom_JCP_64} as:
\begin{eqnarray}
	\delta = 1 + \frac{\gamma}{\beta}
	\label{Widom}
\end{eqnarray}
here, considering the $\beta$ and $\gamma$ values determined from the Fig.~\ref{critical_1}(b), the calculated value of $\delta$ using equation~\ref{Widom} is 3.967(8), which shows good agreement with the critical isotherm analysis. Therefore, it can be concluded that while the extracted critical exponents are consistent across different analytical approaches, they do not adhere to any conventional universality class. Therefore, it is crucial to scrutinize whether the obtained critical exponents can generate the scaling magnetic equation of state \cite{Pramanik_PRB_09}:
\begin{eqnarray}  
	M(H,\epsilon)=|\epsilon|^\beta f_{\pm}\left(\frac{H}{|\epsilon|^{\beta+\gamma}}\right)
	\label{scaling}
\end{eqnarray} 
where f$_+$ and f$_-$ denote the regular functions above and below T$\rm_C$, respectively. For the correct values of critical exponents, the reduced magnetization m=M$|\epsilon|^{-\beta}$ versus reduced field h=H$|\epsilon|^{-(\beta+\gamma)}$ curves should collapse into two branches, corresponding to temperatures above and below the T$_C$. Fig.~\ref{critical_1}(f) illustrates the m versus h plot using the exponents obtained from MAP analysis, showing convergence into two universal curves. This branching is more profoundly observed in the log-log plot [inset of Fig. \ref{critical_1}(f)]. Furthermore, the m$^2$ versus h/m plot in Fig.~\ref{critical_1}(g), also exhibits similar branching in the data points, hence indicating that the magnetic interactions are properly re-normalized in the critical regime following a scaling equation of state \cite{GuptJAP24}.

\begin{table*}
	\caption{Critical exponents ($\alpha,\beta,\gamma$ and $\delta$) determined using modified Arrott plot, Kouvel-Fisher approach and critical isotherm analysis for Co$_2$TiSi$_{0.5}$Al$_{0.5}$ and other systems. 
	}
	
	\begin{tabular}{p{3.4cm}p{2cm}p{3.5cm}p{2cm}p{2cm}p{2cm}p{2cm}}
		\hline
		\hline
		Sample/model & Reference & Approach & $\alpha$ & $\beta$ & $\gamma$ & $\delta$ \\
		\hline
		Co$_{2}$TiSi$_{0.5}$Al$_{0.5}$& This work & Modified Arrott plot &  & 0.369(1) & 1.095(8) & 3.967   \\
		& & Kouvel-Fisher &  & 0.361(5) & 1.108(6) &  4.069 \\	
		& & Critical Isotherm &  &  &  & 3.943(4)  \\		
		Co$_{2}$TiSi$_{0.75}$Al$_{0.25}$& \cite{Datta_PSSB_2020}& Modified Arrott plot &  & 0.429(5) & 1.211(5) & 3.82(6)  \\
		& & Kouvel-Fisher &  & 0.469(1) & 1.173(1) & 3.501(4) \\
	    & & Critical Isotherm &  &  &  & 3.100(2)  \\
	    Co$_{50}$Cr$_{25}$Al$_{25}$& \cite{Panda_JALCOM_2015}& Modified Arrott plot &  & 0.488(3) & 1.144(4) & 3.336(5)  \\
	    & & Kouvel-Fisher &  & 0.482(13) & 1.148(16) & 3.382(20) \\
	    & & Critical Isotherm &  &  &  & 3.401(4)  \\
		Co$_{2}$Cr$_{0.75}$Ti$_{0.25}$Al& \cite{Nehla_PRB_19}& Modified Arrott plot &  & 0.496(4) & 1.348(12) & 3.82(6)  \\
 		& & Kouvel-Fisher &  & 0.491(9) & 1.327(13) &  \\
		& & Critical Isotherm &  &  &  & 3.581(5)  \\
		Co$_{2}$TiGe& \cite{Mandal_PRB_19}& Modified Arrott plot &  & 0.495 & 1.325 & 3.677  \\
		& & Kouvel-Fisher &  & 0.495(2) & 1.324(4) & 3.675  \\
		& & Critical Isotherm &  &  &  & 3.671(1)  \\
	    
	    Co$_{2}$TiSn& \cite{Rahman_PRB_19}& Modified Arrott plot &  & 0.527(3) & 1.229(2) & 3.33(2)  \\
	    & & Kouvel-Fisher &  & 0.537(2) & 1.255(3) & 3.33(7)  \\
	    & & Critical Isotherm &  &  &  & 3.261(2)  \\
	    Co$_{2}$HfSn& \cite{Rahman_PRB_21}& Modified Arrott plot &  & 0.472(3) & 1.021(2) & 3.16(2)  \\
	    & & Kouvel-Fisher &  & 0.477(4) & 1.029(5) & 3.15(7)  \\
	    & & Critical Isotherm &  &  &  & 3.273(5)  \\

		Mean-field& \cite{Kaul_JMMM_85}  & Theoretical & 0 & 0.5 & 1.0 & 3  \\
		3D Heisenberg & \cite{Kaul_JMMM_85}  & Theoretical & -0.115 & 0.365 & 1.386 & 4.80  \\
		3D Ising & \cite{Kaul_JMMM_85} & Theoretical & 0.11 & 0.325 & 1.241 & 4.82 \\
		Tricritical mean-field & \cite{Kim_PRL_02} & Theoretical &  & 0.25 & 1.0 & 5  \\
			
		\hline
		\hline
	\end{tabular}
\label{Tt}
\end{table*}

Furthermore, the nature and range of complex spin interactions play a pivotal role in determining the magnetic properties of a system, which necessitates their investigation. According to the renormalization group theory of exchange-interaction systems, for a $d$-dimensional spin system with an isotropic $n$-component order parameter, the isotropic interaction decays as \( J(r) \sim r^{-(d+\sigma)} \), where \( r \), \( d \), and \(\sigma\) represent the spatial distance, lattice dimensionality, and range of magnetic interactions, respectively. This model classifies the magnetic interaction as long-ranged for \(\sigma\) $\le$ 1.5, and short-ranged for \(\sigma\) $\ge$ 2 \cite{Tarhouni_RSC_2018}. The value of \(\sigma\) can hence be calculated by substituting \(\gamma\) in the following relation \cite{Fisher_PRL_72, Fisher_RMP74, Fischer_PRB_02}
\begin{equation}
	\begin{aligned}
		\gamma= & 1+\frac{4}{d}\left(\frac{n+2}{n+8}\right) \Delta \sigma+\frac{8(n+2)(n-4)}{d^2(n+8)^2} \\
		& \times\left\{1+\frac{2 G\left(\frac{d}{2}\right)(7 n+20)}{(n-4)(n+8)}\right\} \Delta \sigma^2
	\end{aligned}
\end{equation}
where $\Delta \sigma = (\sigma - \frac{d}{2})$, $G(\frac{d}{2}) = 3 - \frac{1}{4} (\frac{d}{2})^2$, and $n$ represents the spin dimensionality. By considering the three-dimensional lattice and spin interactions ($n = d =$ 3) and using the values of $\gamma$ extracted from the MAP, we find $\sigma =$ 1.696, which falls between the mean-field theory and 3D Heisenberg model range, indicating extended long-range magnetic interactions \cite{Tarhouni_RSC_2018, Rahman_PRB_19}. The extracted values of $\sigma$ can further be used to calculate the critical exponent $\alpha$ using the relation $\alpha = 2 - \nu d$, where $\nu$ is the exponent of correlation length, given as $\nu = \gamma / \sigma$. On the other hand, the calculated value of $\alpha$ (0.063) suggests the predominance of long range interactions (mean-field theory, i.e., $\alpha \rightarrow 0$) in the present sample. The magnetic interactions in this sample decay approximately as $J(r) \sim r^{-4.6}$, suggesting that they do not completely adhere to any theoretical model, but lies within long range (mean-field model) and short range (3D Heisenberg model) FM interactions \cite{Rahman_PRB_19, Nehla_JAP_2019}.

\subsection{First-principles simulations}

Finally, to support our experimental results we perform DFT based first-principles calculations to investigate the electronic structure and magnetic properties. The properties in a solid are strongly governed by its crystal structure 
and electronic bonding \cite{Baral_PRB_2022,Klaer_PRB_2009}. Therefore, to 
verify the accuracy of our DFT parameters, we first performed electronic structure calculations for pristine Co$_2$TiSi (CTS) system, presented in Fig.~3 of \cite{SI}. The calculations predict half-metallic character, with 100\% spin polarization (SP) and a band gap of 0.60 eV associated with minority spin channel, concurring with reported literature \cite{Kandpal_JP_2007, Sharma_JMMM_2010}. The majority spin channel exhibits metallic character due to the Co$-$Ti interactions, while the Co$-$Co second next neighbor direct hybridization is responsible for semiconducting nature of minority spin channel \cite{Okutani_JPSJ_2000, Aguayo_JMMM_2011}. The computed lattice parameter is 5.7475 \AA, which agrees well with the calculated value of 5.760 $\rm \AA$ in \cite{Kandpal_JP_2007} and experimentally observed value of 5.743~\AA~in \cite{Carbonari_HI_1993}. Additionally, as evident from the Table \ref{T_lp}, the computed magnetic moment, 1.99 $\mu \rm _B$/f.u, is in very good agreement with refs.~\cite{Kandpal_JPDAP_2007,Sharma_JMMM_2010},  which substantiates the exchange-correlation functional and other 
 parameters used in the calculations.

\begin{figure*}
\includegraphics[width=7.2in]{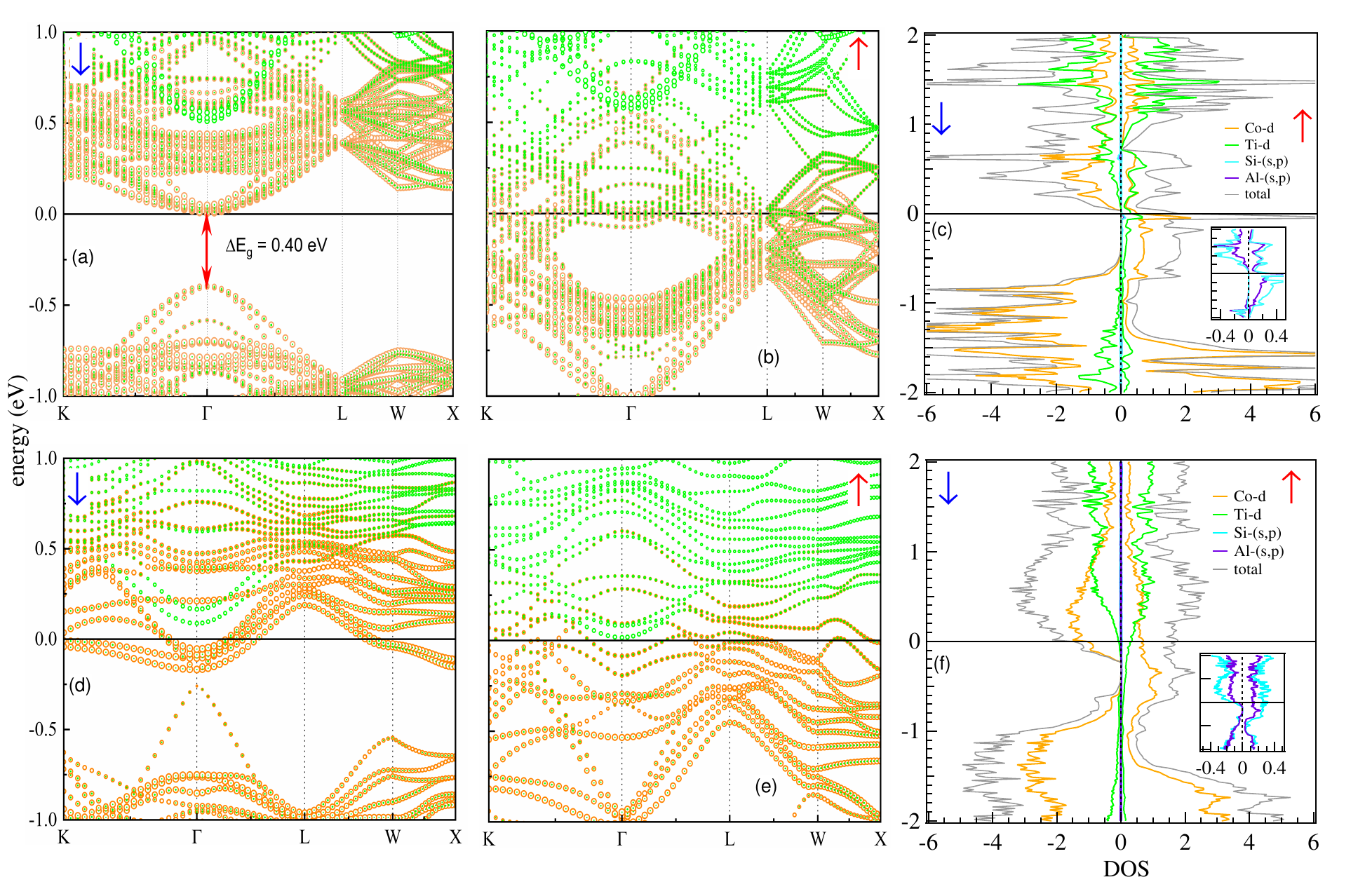}
\caption {(a, b) The atom-resolved electronic band structure of Co$_2$TiSi$_{0.5}$Al$_{0.5}$ (CTSA), near Fermi level for minority and majority spin states, respectively. (c) The total and partial density of 
	states of majority and minority spin channels are shown corresponding to 
	near Fermi level region for the CTSA system. (d, e) The atom-resolved electronic band structure of with disorder, i.e., CTSA(D) system, near Fermi level for minority and majority spin states, respectively. (f) The total and partial density of 
	states of majority and minority spin channels in CTSA(D) system.}
 \label{bands}
\end{figure*}

\begin{table}[h]
	\caption{Total magnetic moment and atom resolved magnetic moment per formula unit. 
	}
\begin{tabular}{p{2.7cm}p{1cm}p{1.1cm}p{1.1cm}p{1.1cm}p{1.1cm}}
\hline
\hline
System & $m^{\rm Co}$ & $m^{\rm Ti}$ &$m^{\rm Si}$ &$m^{\rm Al}$ &$m^{\rm total}$ \\
& ($\mu \rm _B$) & ($\mu \rm _B$) & ($\mu \rm _B$) & ($\mu \rm _B$) & ($\mu \rm _B$) \\
\hline
 	 Co$_{2}$TiSi& 0.99 & -0.04 & 0.03 & - & 1.99  \\
     
      Co$_{2}$TiSi \cite{Kandpal_JP_2007}& 1.03 & -0.02 & - & - & 2.00  \\
      Co$_{2}$TiSi \cite{Sharma_JMMM_2010}& 0.99 & 0.03 & 0.01 & - & 1.99  \\
      \\ 
      Co$_{2}$TiSi$_{0.5}$Al$_{0.5}$& 0.822 & -0.097 & 0.011 & 0.001 & 1.559  \\ 
      Co$_{2}$TiSi$_{0.5}$Al$_{0.5}(\rm D)$& 0.580 & -0.012 & 0.008 & -0.003 & 1.160  \\  
      Co$_{4}$Ti$_{2}$SiAl \cite{Zareii_PBCM_2012} & 0.844 & -0.111 & 0.015 & -0.013 &   \\  
   
\hline
 \hline
\end{tabular}
\label{T_lp}
\end{table}

Further, using the optimized unit cell of pristine CTS sample, the Co$_{2}$TiSi$_{0.5}$Al$_{0.5}$ system was constructed by creating a 2$\times$2$\times$2 supercell followed by substituting 50\% Si by the Al atoms, see Fig.~\ref{XRD}(b). The substituted structure is further optimized to obtain the global minimum state, where the optimized lattice parameters are obtained as $a = b = 5.7802$ \AA~ and $c= 5.8003$ \AA, which are consistent with the value of 5.8047 \AA~in ref.~\cite{Zareii_PBCM_2012}; however, with a slight distortion along the 'c' axis. The calculated magnetic moment 1.56 $\mu \rm _B$/f.u. is found to be close to the Slater-Pauling value, 1.5 $\mu \rm_B$. Also, the calculated total and atom resolved magnetic moments of present sample are in very good agreement with reported values in \cite{Zareii_PBCM_2012}, as summarized in Table \ref{T_lp}. However, there is a significant difference from the experimentally observed magnetic moment value, 1.2 $\mu \rm_B$/f.u in Fig.~\ref{MT_100}(b), which could possibly be due to the presence of chemical disorder between the constituents Y and Z elements \cite{Jezierski_JMMM_1996}. Therefore, we examine the impact of anti-site disorder through swapping of sites between Ti and Al atoms by constructing all the possible structures corresponding to a particular disorder level followed by selection of the most energetically favorable structure for subsequent calculations. Interesting, we find a systematic decrease in the total magnetic moment with increasing level of disorder,  as summarized in Table~I of \cite{SI}, owing to the corresponding change in electronic structure. The system corresponding to 37.5\% disorder-level Co$_2$TiSi$_{0.5}$Al$_{0.5}$(D), renders the calculated total magnetic moment of 1.16 $\mu \rm_B$/f.u., which is in excellent agreement with the experimental value (1.2 $\mu \rm_B$/f.u.). Furthermore, using the theoretical value of total magnetic moment in 
equation $T_{\rm C} = 23 + 181*m^{\rm total}$ {\cite{Jin_AIP_2017}, 
the extracted $T_{\rm C}$ value is 233 K, which underestimates the experimental value of 278 K. This can be attributed to the empirical nature of this relation, as the structural disorder significantly alters the electronic structure and hence the magnetic correlations in the system. As evident in the Table \ref{T_lp}, the atom-resolved calculations reveal that Co atoms have the dominant magnetic moment of 0.58 $\mu \rm_B$/f.u., followed by a considerable antiparallel contribution of -0.012 $\mu \rm_B$/f.u. 
from Ti, and small contributions of $\approx$ 0.008 and -0.003 $\mu \rm_B$/f.u. 
from Si and Al atoms, respectively. The hybridization of Co-{\it d} states with Ti-{\it d} states, hence induces a small magnetic moment at Ti site, even though in the opposite direction.

The electronic structure of Co$_2$TiSi$_{0.5}$Al$_{0.5}$ sample is presented in 
Fig.~\ref{bands} showing both atom-projected electronic band structure and density of states (DOS) for  CTSA (top panel) and with disorder CTSA(D) (bottom panel). The majority spin channel shows `metallic' characteristics, whereas there is a finite band gap of 0.40 eV in the minority spin states confirming the persistence of half-metallic character, as discernible from the panels (a) and (b), on Al-substitution. From the PDOS for majority spin channel, panel (c), the valence and conduction bands are densely populated with Co-$d$ and Ti-$d$ electrons, with nonzero contributions from $sp$-elements Si and Al. However, for the minority channel, there are no states in the vicinity of the Fermi level in the valence band, whereas dominant contributions from Co-$d$ and Ti-$d$ electrons are observed for the conduction band. This is in contrast with a previous study, where around 50\% Fe doping in Co$_2$TiSi was reported to exhibit loss of half-metallicity owing to the emergence of minority states at the Fermi level \cite{Jin_AIP_2017}. The spin polarization calculated from the total density of states for the Co$_2$TiSi$_{0.5}$Al$_{0.5}$ sample, panel (c), renders the 
value 100\%. This implies that upon Al substitution, even-though the band splitting in minority spin channel reduces from 0.60 eV to 0.40 eV; the half-metallic nature is still preserved. On the other hand, when the Ti-Al disorder is introduced, we observe a significant decrease in the SP value from 100\% for CTSA to 8.3\% for CTSA(D), symbolizing the loss of half-metallicity from the calculated bands and DOS, as shown in panels (d, e, f). The plausible cause of this drastic reduction of SP and loss of half-metallic nature could be associated to the emergence of Co-$d$ states in the vicinity of Fermi level for minority channel, in the disordered system. Here, we also find that, similar to the undistorted structure, the valence and conduction bands of majority states have dominant contributions from Co-$d$ and Ti-$d$ electrons. However, the minority channel are now composed of dominating Co-$d$ electrons as well as some nonzero Ti-$d$ contribution at the Fermi level. This study therefore highlights the profound influence of structural disorder on the electronic structure of the system, which could hence be used to tune the magnetic and transport properties of prospective materials for future spintronic applications \cite{Kelvin_STAM_2021}. 

\section{\noindent ~Conclusions}

In conclusion, our comprehensive study of the Co$_2$TiSi$_{0.5}$Al$_{0.5}$ system reveals a complex interplay between structural disorder and observed magnetic properties. Through a combination of experimental measurements and theoretical simulations, we have demonstrated that the atomic disorder significantly impacts the underlying magnetic interactions, subsequently leading to the emergence of a magnetically in-homogeneous Griffiths-like phase. The Ti-Al anti-site disorder results in random allocation of different exchange-coupling constants between Co-(Ti/Al)-Co atoms, hence manifesting multiple magnetic transitions in the system. The profound field and temperature dependence of these magnetic correlations was observed in the lower magnetic fields (H$\leqslant$ 0.1 Tesla) entropy curves, exhibiting multiple peaks corresponding to these transitions. Furthermore, the calculated values of critical exponents $\alpha, \beta, \gamma$ and $\delta$ are $0.063, 0.361, 1.108$ and $3.943$, respectively, which are slightly away from mean-field theory and deviating towards Heisenberg model. However, the corresponding magnetic interactions decay approximately as $J(r) \sim r^{-4.6}$, reinforcing the notion of extended long-range magnetic ordering in the sample. The persistence of half-metallicity with Al substitution, albeit disrupted by structural disorder, underscores the role of electronic structure in determining the magnetic behavior of the system. The disorder significantly alters the electronic structure of the system, evident from the reduced magnetic moment 1.16 $\mu \rm_B$/f.u and spin polarization 8.3\%, due to emergence of finite Co-$d$ states at Fermi level in the minority spin channel. Our findings therefore contribute to a deeper understanding of the magnetic interactions in Co-based Heusler alloys, paving the way for future explorations of their practical applications in spintronics and magnetic refrigerators. 

\section{\noindent ~Acknowledgments}
PY thanks MHRD for the fellowship and expresses sincere gratitude towards Dr. Ajay Kumar, Dr. Zeeshan and Madhav for useful discussions. We acknowledge the Department of Physics, IIT Delhi for providing XRD and MPMS facilities and central research facility (CRF) for FE-SEM measurements. The theoretical results presented in this work have been computed using the High Performance Computing cluster Padum, at IIT Delhi. RSD acknowledges the BRNS for financial support through DAE Young Scientist Research Award with Project Sanction No. 34/20/12/2015/BRNS..

\end{document}